\documentclass[aps,pre,epsf,twocolumn,floatfix]{revtex4-2}
\usepackage{amsmath}
\usepackage{amssymb}
\usepackage{epsfig}
\usepackage{graphicx}
\usepackage[T1]{fontenc}
\usepackage[utf8]{inputenc}
\usepackage{color}
\usepackage[normalem]{ulem}
\usepackage{tikz}
\usetikzlibrary{positioning}

\begin{document}

\title{Crossover of Critical Behavior in Dynamic Phase Transitions of Multilayer Ising Model Systems}

\author{Erol Vatansever}
\email{erol.vatansever@deu.edu.tr}

\affiliation{Department of Physics, Dokuz Eyl\"{u}l
University, TR-35160, Izmir-Turkey}

\author{Mikel Quintana}
\email{mikel.quintanauriarte@tecnalia.com}

\author{Andreas Berger}
\email{a.berger@nanogune.eu}

\affiliation{CIC nanoGUNE BRTA, E-20018 Donostia-San Sebastian, Spain}

\date{\today}

\begin{abstract}
We investigate the crossover of critical behavior for the dynamic phase transition (DPT) in ferromagnetic thin films using Monte Carlo simulations of the kinetic Ising model, focusing on the scaling behavior of the dynamic order parameter under a time-dependent external magnetic field.  Specifically, we study the transition of the critical behavior of such multilayer film systems from two-dimensional (2D) to three-dimensional (3D) as a function of the film thickness and the distance to the critical point, which enables dimensional crossover observations. Our results indicate that the effective critical exponents exhibit a clear transition in their scaling behavior, with thinner films showing 2D-like characteristics and thicker films displaying 3D-like behavior, for both the DPT and the thermodynamic phase transitions (TPT). Quantitatively, the crossover from 2D to 3D behavior occurs at larger film thicknesses for the DPT compared to the TPT, suggesting that DPT and TPT are governed by distinctly different length scales and underlying surface effects. These findings are in agreement with experimental observations in ultrathin Co films, where dynamic and thermodynamic critical exponents were found to differ. Therefore, our study provides an in-depth explanation for critical phenomena in thin-film ferromagnets driven by a time-dependent magnetic field. By comparing the dimensional crossover properties of both TPT and DPT, we present
a comprehensive understanding of how thin-film geometry and surface effects influence the scaling laws and critical behavior in nonequilibrium systems. 
\end{abstract}

\maketitle

\section{Introduction}
\label{sec:intro}
Phase transitions are one of the most intriguing phenomena in many-body physics, describing an abrupt change of macroscopic observables in a system upon slightly changing an external control parameter, such as the temperature, for instance, \cite{Goldenfeld}. Conventionally, physics research has been focused on TPT, which occur under thermodynamic equilibrium conditions, but phase transitions are also observed in systems that are out-of-equilibrium \cite{Henkel}. Relevant examples include superconducting materials \cite{Benkraouda}, charge density waves \cite{Ogawa, Bonamy}, or laser emission \cite{Kröll}. Such non-equilibrium phase transitions as well as DPTs are, therefore, an extremely important phenomenon in the context of non-equilibrium physics in many different types of systems.

Following the need to understand such non-equilibrium processes, systems that can exhibit both thermodynamic equilibrium and dynamic phase transitions are particularly relevant. This is because they allow us to compare the occurrences, similarities, and differences between both the TPT and DPT \cite{Campajola, Vajna}. Hereby, ferromagnets represent very relevant example systems exhibiting both types of phase transitions \cite{Riego}. Furthermore, their similarities, particularly in the context of the corresponding scaling behavior and universality, make them a prominent system for the study of dynamic non-equilibrium processes in general.

The DPT of ferromagnets was originally studied in the context of the mean-field approximation (MFA) of the kinetic Ising model (KIM) \cite{Tome}. Within the KIM, the dynamic behavior of a model ferromagnet is described by a Hamiltonian with spin-1/2 as,

\begin{equation}\label{Hamiltonian}
    \mathcal{H} = - J \sum_{\{ ij\}}S_iS_j-H(t)\sum_{i}S_i,
\end{equation}
with $S_i = \pm 1$ representing the spin states on a crystal lattice, $J$ being the next-neighbor exchange coupling strength and $H(t)$ being the external time-dependent magnetic field. For the past three decades, many theoretical studies in the context of the MFA \cite{Idigoras, Fujisaka, Keskin1, Keskin2, Keskin3} and Monte Carlo simulations \cite{Rikvold, Sides, Park1, Park2, Taucher, Buendia1, Buendia2, Buendia3} have elaborated an understanding of the DPT, particularly in the context of the Ising model. More recently, several experimental works have observed a number of unexplored phenomena and verified crucial aspects of the DPT that were previously predicted \cite{Berger, Riego2, Marin, Quintana}.

More specifically, the DPT of ferromagnets describes the abrupt change of time-dependent magnetization trajectories $M(t)$ upon changing the parameters of a periodic time-dependent external magnetic field $H(t)$ of period $P$, at temperatures $T$ below the Curie temperature $T_C$. Such an abrupt change in $M(t)$ occurs when $P$ becomes comparable to the relaxation time-constant $\tau$ of the spin system. Hereby, the period-averaged magnetization $Q$, calculated as,

\begin{equation}\label{OParameter}
    Q = \frac{1}{P}\int_{t}^{t+P}M(t')dt'
\end{equation}
represents the order parameter associated with the DPT. For a fixed oscillating field amplitude $H_0$, one observes $Q\neq0$ below some critical period $P_c$, which corresponds to the occurrence of a dynamically ordered or dynamic ferromagnetic phase, while for $P > P_c$, $Q = 0$, which corresponds to the dynamically disordered or dynamic paramagnetic phase. Thus, $P_c$ represents the critical point at which the system undergoes a second-order phase transition (SOPT).

A constant bias field $H_b$, superimposed to the field oscillations, was shown to play the role of the conjugate field of $Q$ \cite{Berger, Robb} for simple field sequences such as sinusoidal or square type \cite{Robb2, Quintana2}, identifying the $(P, H_b)$ plane as the DPT's  natural phase space in which all key characteristics of the DPT can be observed. Upon switching $H_b$ in the dynamically ferromagnetic phase, the system will undergo a first-order phase transition (FOPT), inverting the sign of the order parameter $Q$. This FOPT can also lead to the observation of a metastability regime and subsequent hysteretic behavior of $Q(H_b)$ in the dynamically ferromagnetic phase \cite{Berger}. These observations are formally identical to the ones in the $(T, H)$ phase space of the TPT, for which $M$ is the equilibrium order parameter \cite{Riego}. 

Another key feature occurring in both the TPT and DPT is the scaling behavior of the respective order parameter in the vicinity of the SOPT. Indeed, for $H_b = 0$, it is known that,

\begin{equation}
\label{Scaling1}
    Q(P\rightarrow P_c, H_b = 0) \propto (P_c-P)^{\tilde{\beta}},
\end{equation}
where $\tilde{\beta}$ is a dynamic critical exponent. Similarly, for $P = P_c$ it is known that,

\begin{equation}
\label{Scaling2}
Q(P = P_c, H_b \rightarrow 0) \propto H_b^{1/\tilde{\delta}},
\end{equation}
with $\tilde{\delta}$ being another relevant dynamic critical exponent. The scaling behavior in Eqs. (\ref{Scaling1}), and (\ref{Scaling2}) are formally identical to those of the magnetization $M$ in the thermodynamic equilibrium phase space, with

\begin{equation}
\label{Scaling3}
    M(T\rightarrow T_C, H = 0) \propto (T_C-T)^{\beta},
\end{equation}

\begin{equation}
\label{Scaling4}
M(T = T_C, H \rightarrow 0) \propto H^{1/\delta},
\end{equation}
with $\beta$ and $\delta$ being thermodynamic equilibrium critical exponents \cite{Arrott}. Furthermore, both phase transitions have been shown to correspond to the same universality class and, consequently, the critical exponents are identical for systems with the same dimensionality \cite{Fujisaka}.

Both, the scaling behavior and universality of the DPT have been extensively documented in different theoretical works \cite{Sides2, Korniss}. More recently, the dynamic critical exponents of the 2D Ising model have been observed experimentally in ultrathin Co films \cite{Quintana3}. However, this experimental study also led to another interesting observation, namely, that the thermodynamic equilibrium critical exponents $\beta$ and $\delta$ agreed with those of the 3D Ising model, while $\tilde{\beta}$ and $\tilde{\delta}$ agreed with those of the 2D Ising model in the same film. 

To explain this intriguing result and seeming contradiction, the authors of \cite{Quintana3} noted that their films were several monolayers thick and thus represented a multilayer thickness regime, in which dimensional crossover from 2D to 3D type behavior had been observed for the TPT, both theoretically and experimentally \cite{Li}. So, if TPT and DPT exhibit a difference in the thickness range, at which this dimensional crossover takes place, the seemingly contradictory experimental results could be explained. This implies that upon using the same relative critical regime criteria, the characteristic length-scale, at which the critical exponents change from the 2D to the 3D case would be different in the TPT and DPT. Accordingly, a single film of finite thickness could exhibit different exponents for both types of phase transitions. 

In this regard, it is noteworthy to mention that non-universal DPT behavior has already been observed at the surfaces of systems, for which the DPT can be entirely absent, even if it occurs in the associated bulk system \cite{Park, Riego3}. Thus, it is possible that the dimensional crossover is similarly affected by the existence of two surfaces in a finite thickness film, leading to a different length-scale for TPT and DPT, at which dimensional crossover occurs, even in the exact same physical system.  To the best of our knowledge, the aspect of dimensional crossover for the DPT has not been studied to date \cite{Vatansever, Chakrabarti, Acharyya}, and so, it will be the focus of our study here.
In this work, we explore by means of extensive Monte Carlo simulations of the Ising model the critical scaling of the DPT for model films with different thicknesses and we compare their scaling behavior with those of the TPT. In the second section of this work, we explain the key aspects of our model and Monte Carlo calculations. In the third section, we explain the results of our work and in the fourth section, we draw some general conclusions and provide an overall outlook.

\section{Simulation framework}
\label{sec:framework}
We simulate the ferromagnetic thin film system described by the Hamiltonian in Eq. (\ref{Hamiltonian}) where the external magnetic field has a square-wave profile with an amplitude $H_0$ and a period $P$.  In our Monte Carlo simulations, the boundary conditions are chosen to reflect the physical constraints of real thin-film systems, as shown schematically in Fig. (\ref{System}). Specifically, periodic boundary conditions are imposed in the $x$ and $y$ directions to mimic a laterally infinite film system and minimize finite-size effects. Free boundary conditions are employed along the $z$ direction, allowing for the consideration of surface effects that are essential in modeling the thin-film geometry.  The system we study consists of $N = L_x \times L_y \times  L_z $ spins, where $L_x$ and $L_y$ denote the linear sizes of the system in the $x$ and $y$ directions while $L_z$ represents the number of layers and thus the thickness of the multilayer film along the $z$ direction. 
During the simulations, the length of the $x$ edge of the thin films will be chosen equal to the length of the $y$ edge, and they will be represented as $L_x =L_y = L$. Such kind of geometry is convenient for modeling thin films of varying thicknesses and lateral dimensions and makes it possible to analyze size-dependent properties, surface effects, and dimensionality-driven phenomena. We note that our data evaluations are conducted across 100 independent simulation runs, and to compute error bars, we use the Jackknife method \cite{Newman} for both TPT and DPT. Since we aim to compare the dimensional crossover phenomena appearing in TPT and DPT, the simulation frameworks for these two different cases will be presented in separate subsections.

\begin{figure}[h!]
\label{System}
\begin{center}
\includegraphics[scale=0.4]{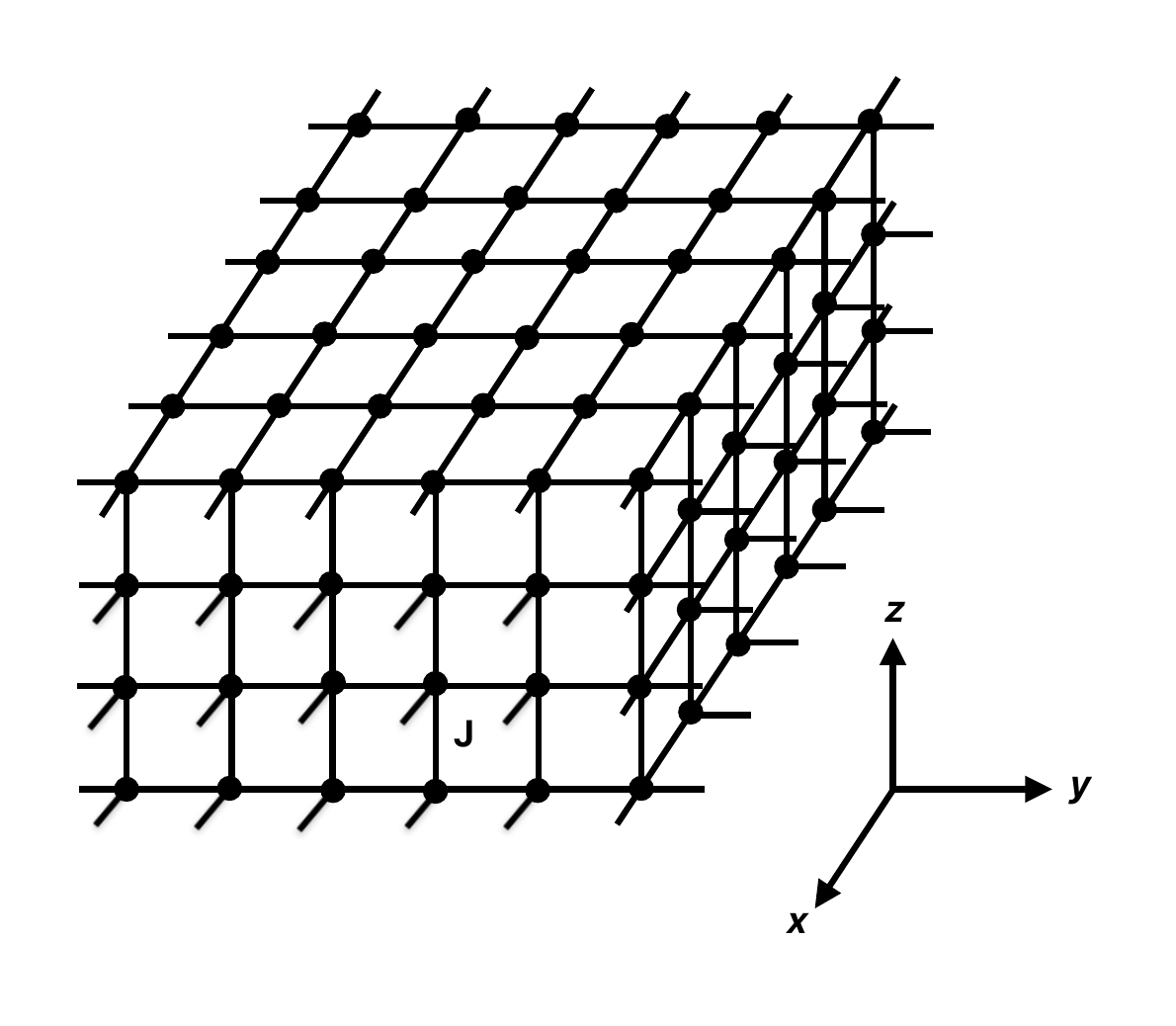}
\caption{Schematic of a $L_z = 4$ thin film system displaying our selected surface orientation and interaction structure. Surface atoms are connected to four in-plane nearest neighbors and one in the adjacent layer, while bulk atoms are symmetrically coupled to six nearest neighbors. All spin-spin interactions are ferromagnetic, and they are described by a uniform coupling constant $J$.}\label{System}
\end{center}
\end{figure}

\begin{figure}[h!]
\begin{center}
\includegraphics[scale=1.0]{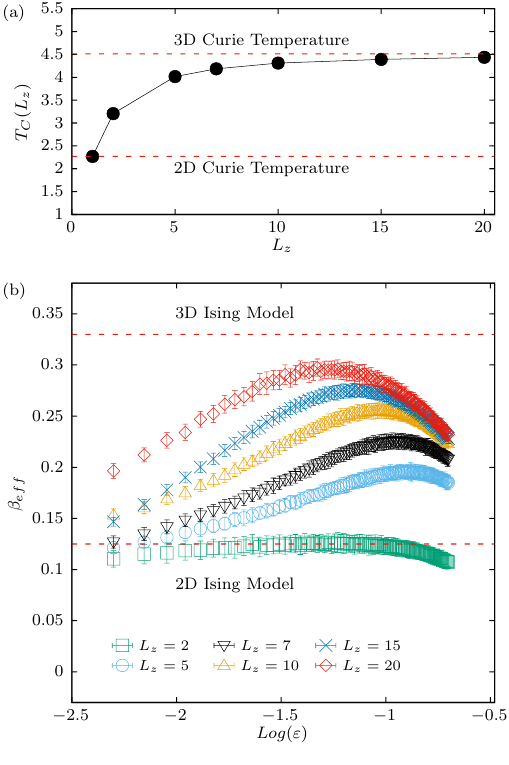}
\caption{(a) Film thickness $(L_z)$ dependence of the Curie temperature $T_C$ of the thermodynamic phase transition, reflecting the shift from two-dimensional to bulk-like behavior. (b) Effective critical exponent $\beta_{eff}$ obtained from Eq. (\ref{beta_TPT}) for film thickness from $L_z = 2$ to $20$, providing insights into the critical properties of a finite thickness system in equilibrium. For the Ising model, the Curie temperatures (a) and critical exponents (b) for both two-dimensional and three-dimensional systems are indicated by red dashed lines.}\label{Fig2}
\end{center}
\end{figure}

\subsection{Simulation protocols and extracted quantities for TPT analysis}
\label{sec:model1}
We begin by presenting the computational framework designed for TPTs. To elucidate the equilibrium critical behavior of the system, we eliminate the influence of the external magnetic field by setting the reduced field amplitude to zero $(H_0/J = 0)$.  To overcome the critical slowing-down and obtain a substantial improvement in simulation performance near the phase transition, we utilize the Swendsen-Wang cluster algorithm \cite{Newman, Landau}. We measure several key quantities to characterize the behavior of the system near the critical point starting from the total magnetization per site:

\begin{equation}\label{TotalMag}
M  =\frac{1}{N} \sum_{i=1}^N S_i.   
\end{equation}

Since the magnetization can take both positive and negative values in the absence of an external magnetic field, we evaluate the absolute value of $M$ when calculating the order parameter.

To investigate the evolution of the effective critical exponents as the system transitions from a thin film to bulk behavior, it is essential to first analyze the variation of the Curie temperature with the thickness of the system.  For a range of film thicknesses, the Curie temperatures are computed using the fourth-order Binder cumulant $U_L$ method \cite{Newman, Binder}, which is a well-established approach for locating Curie temperatures with high precision, utilizing the definition:
\begin{equation}\label{Binder}
   U_L = 1-\frac{\langle M ^4\rangle}{3\langle M ^2 \rangle^2}, 
\end{equation}
where $\langle\cdots\rangle$ denotes the thermal average. Each simulation consists of $5\times10^4$ Monte Carlo steps (MCSs) per temperature, providing sufficient data for a meaningful analysis. To guarantee that the system reaches equilibrium before data are collected, the first $10^4$ of the MCSs are discarded to facilitate the thermalization process. Simulations are conducted for system sizes of $L = 256$ and $L = 512$, with calculations performed for various values of the film thickness $L_z$ which allows for a detailed examination of the interplay between system size and critical behavior. Numerical results for the Curie temperature $T_C$ as a function of $L_z$ are depicted in Fig. \ref{Fig2}(a). Here, the Curie temperatures for 2D and 3D Ising models with periodic boundary conditions in all directions are denoted by red dashed lines \cite{Onsager, Ferrenberg}.  It is clear from the figure, that upon increasing the thickness $L_z$ starting from a 2D monolayer $(L_z=1)$, the Curie temperature rises steeply driven by the enhanced stability of the system due to effectively stronger interactions and an accordingly reduced impact of thermal fluctuations. We can also see that even for thicknesses $L_z$ of less than 10 layers, the Curie temperature approaches the 3D Ising case already, which implies that also the critical behavior and exponents will be relevantly impacted, even if a finite thickness film is formally a 2D system. At this point, we note that the numerical outcomes reported here for the Curie temperatures are 
in good agreement with previous studies \cite{Binder2, Schilbe, Marquez}. 

After determining the Curie temperatures for each $L_z$ value, we calculate  the effective critical exponent $(\beta_{eff})$ defined as: 
\begin{equation}\label{beta_TPT}
\beta_{eff} = \frac{\partial\;\text{Log}M}{\partial\; \text{Log}\varepsilon}\Bigr\rvert_{H = 0}, \end{equation}
where $\varepsilon$ denotes the temperature deviation relative to $T_C$ for each $L_z$ value that can be represented by the following equation:
\begin{equation}\label{eps}
\varepsilon = \frac{T_C(L_z)-T}{T_C(L_z)}.    
\end{equation}

\noindent Here, $T$ is the temperature of the system. The corresponding results are shown in Fig. \ref{Fig2}(b) and will be discussed in Sec. \ref{sec:results}.

\subsection{Simulation protocols and extracted quantities for DPT analysis}
\label{sec:model2}
To study the dynamic phase transitions, we carry out Monte Carlo simulations on the thin film system using the single-site update Metropolis algorithm \cite{Newman, Metropolis}. To enhance the efficiency of the simulations, we implement a geometric parallelization scheme, in which the lattice is divided into multiple strips. The division is based on both the thickness of the system and the number of available processors such that the computational load is distributed optimally across the parallelized system. The protocol during the simulations is as follows: We discard the first $10^3$ periods of the external field to allow the system to thermalize. Numerical data are then collected and analyzed during the subsequent $10^4$ periods of the external field cycle. As mentioned above, a square-type magnetic field has been selected as the source of the time-dependent magnetic field due to its lower computational cost \cite{Vatansever}. During the simulations, the times and, consequently, the period of the external magnetic field are expressed in terms of MC steps. The time, at which the system's initial configuration is prepared is selected as the initial time. When the Metropolis algorithm is applied to one sweep of the lattice, an MC step is performed during which the external magnetic field is fixed. Specifically, the field value is held constant for $P_{1/2}$ MC steps, after which it is inverted. Here, $P_{1/2}$ is the half-period of the external magnetic field. In the new MC step, all spins in the system experience the corresponding value of the external magnetic field at that moment, and this process continues in the same manner. Therefore, as the simulation progresses, the system experiences the effect of an external magnetic field that varies periodically with an amplitude between the values of $H_0$ and $-H_0$.

At this point, it is crucial to emphasize that DPT occurs within the multidroplet (MD) regime, where metastable decay is driven by the nucleation and subsequent growth of numerous droplets. This regime is characterized by the transition of the system from a metastable state to a stable state through the collective behavior of many droplets. To study the MD regime, the field amplitude is fixed at $H_0/J = 0.4$ in this study while the temperature is set to $T = 0.8T_C(L_z)$ where $T_C(L_z)$ denotes the thickness $L_z$ dependent Curie temperature \cite{Vatansever}. We determine the critical half-period, at which dynamic phase transition occurs by utilizing the fourth order Binder cumulant  $\tilde{U}_L$ method \cite{Vatansever}, which is given as:
\begin{equation}\label{BinderDPT}
 \tilde{U}_L = 1- \frac{\langle Q^4\rangle}{3 \langle Q^2 \rangle^2}.   
\end{equation}
For this analysis, we use the system sizes $L = 256$ and $L = 512$, and obtain critical half-period $P_{1/2}^c$ values, which are dependent on the film thickness $L_z$. Once the dynamic critical half-period  values of the system for each $L_z$ value are determined, the effective critical exponent $(\tilde{\beta}_{eff})$ is computed as: 
\begin{equation}\label{beta_DPT}
\tilde{\beta}_{eff} = \frac{\partial\;\text{Log}Q}{\partial\; \text{Log}\tilde{\varepsilon}}\Bigr\rvert_{H_b = 0} \end{equation}
where $Q$ is the dynamic order parameter defined in Eq. (\ref{OParameter}), and as in the thermodynamic case, the dynamic order parameter $Q$ is evaluated using its absolute value to properly account for the symmetry of the system under a periodically changing magnetic field. $\tilde{\varepsilon}$ denotes the half-period deviation relative to $P_{1/2}^c$ for each $L_z$ value, and it can be defined as follows:
\begin{equation}\label{tildeeps}
\tilde{\varepsilon} = \frac{P_{1/2}^c(L_z)-P_{1/2}}{P_{1/2}^c(L_z)},
\end{equation}

\noindent with $P_{1/2}^c(L_z)$ being the critical half-period of the system. We note that although $\varepsilon$ and $\tilde{\varepsilon}$  describe different physical quantities (temperature and half-period, respectively), they both quantify the relative distance from the critical point using the natural control parameter of their respective transitions. Hence, their functional roles are conceptually analogous and allow for meaningful comparison when analyzing the critical behavior and crossover properties of TPT and DPT systems.

As described in Eqs. (\ref {beta_TPT}) and (\ref{beta_DPT}), we employ the direct fitting method to estimate effective critical exponents over a finite temperature (half-period) range near the transition point. While it is well known that this method has limitations in accurately determining asymptotic critical behavior, it remains a widely used approach in the analysis of crossover phenomena, where the focus is on the evolution of effective exponents.  As discussed in Refs. \cite{Schilbe, Marquez}, the direct method offers a straightforward way to capture the temperature-dependent behavior of the order parameter and to identify trends associated with changes in universality class or transition type.

\section{Results and discussion}
\label{sec:results}
In this section, we will discuss the evolution of the effective critical exponents defined in Eqs. (\ref{beta_TPT}) and (\ref{beta_DPT}) as a function of film thickness and distance to the critical point for thermodynamic and dynamic phase transitions. We start this discussion by considering $\beta_{eff}$ as a function of $\varepsilon$ for different values of $L_z$ for the TPT, as displayed in Fig. 2(b). In a general manner, all curves show an $\varepsilon$ dependent $\beta_{eff}$ that exhibits a maximum. For very thin films, the maximum is barely visible, and $\beta_{eff}$ is near its 2D value for almost any $\varepsilon$ value. For thicker films, the peak becomes successively larger, and for the thickest film, it shows a $\beta_{eff}$ peak value approaching the 3D critical exponent. The thicker films first start to approach the 3D critical value upon lowering $\varepsilon$, i.e. in their initial approach towards the Curie temperature, and subsequently invert their trajectory to approach the 2D behaviour at temperatures even closer to $T_C$, given that the correlation length then exceeds the film thickness, leading to the overall peak type behavior as a function of $\varepsilon$. This dimensional crossover phenomenon becomes ever more visible for larger $L_z$ values. More specifically, there are three distinct regions. The first region at larger $\varepsilon$ values corresponds to the temperature range in which the correlation length remains too small to detect the finite thickness of the film. In this region, the correlation length progressively increases upon approaching the critical point, and the effective critical exponents is moving towards the three-dimensional value $(\beta^{3D} = 0.3264)$ \cite{Ferrenberg}. This indicates a transition in the system's behavior as it moves closer to the regime where the finite thickness of the film starts to play a significant role in determining its critical properties. The second region emerges at a temperature where the correlation length becomes comparable to or exceeds the film's finite thickness, thereby revealing the reduced dimensionality of the system. In this regime, the behavior of the system is increasingly governed by its thin-film geometry, which distinguishes it from bulk-like characteristics. The effective critical exponents reach a saturation point and can no longer increase, and this situation marks the onset of a crossover toward the critical exponents characteristic of a two-dimensional system $(\beta^{2D} = 0.125)$ \cite{Zeynep, Fisher, Vasilopoulos}. In the third region, the finite-size effects dominate the critical behavior and make the thin film systems essentially indistinguishable from a pure 2D system.

At this point, we also note that the use of effective critical exponents $\beta_{eff}$, as defined in Eq. (\ref{beta_TPT}), is essential for identifying the $\varepsilon$-range where power-law scaling of the order parameter is valid. While the value at $\varepsilon = 0$ coincides with the actual critical exponent, this value alone is insufficient to characterize the critical region unless the scaling behavior persists across a finite region adjacent to the critical point. In this regard, 
$\beta_{eff}$,  serves as a diagnostic tool for detecting the limits of asymptotic scaling. A constant or nearly constant value of $\beta_{eff}$,
over a finite $\varepsilon$-interval, implies the presence of a well-defined critical region. Conversely, strong variations in $\beta_{eff}$
reflect deviations from this regime, typically associated with crossover effects. As seen in Fig. \ref{Fig2}(b), the $L_z = 2$ case displays a broad region with consistent exponent values, validating its effective two-dimensional nature. For larger 
$L_z$, however, the effective exponent varies significantly with $\varepsilon$, indicating the onset of dimensional crossover. This variation provides a clear quantitative indicator of the crossover phenomenon that is occurring in sufficiently thick films.

\begin{figure}[h!]
\begin{center}
\includegraphics[scale=1.0]{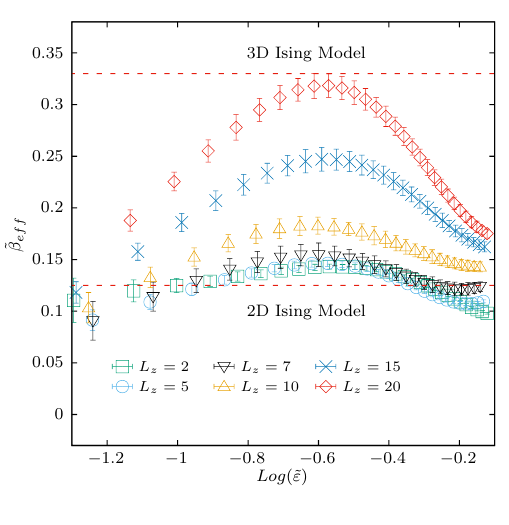}
\caption{
Effective critical exponent $\tilde{\beta}_{eff}$ in the vicinity of the dynamic phase transition point, calculated from Eq. (\ref{beta_DPT}) for varying values of the film thickness from $L_z = 2$ to $20$.  The critical exponents for both two-dimensional and three-dimensional systems are indicated by red dashed lines.}\label{Fig3}
\end{center}
\end{figure}

\begin{figure}[h!]
\begin{center}
\includegraphics[scale=1.0]{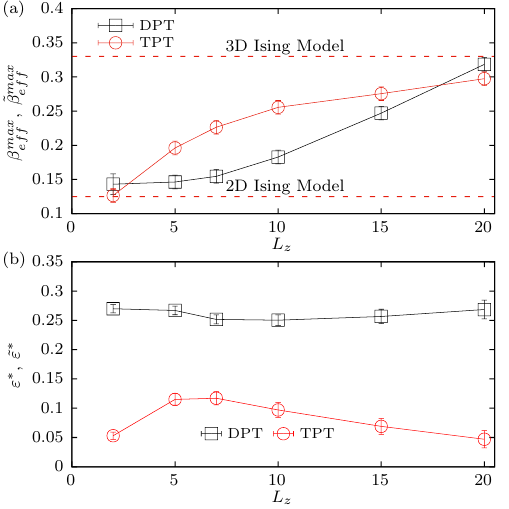}
\caption{(a) Maximum values of the effective critical exponents ($\beta_{eff}^{max}$, $\tilde{\beta}_{eff}^{max}$) as a function of the film thickness, displaying the distinct dimensional crossover behaviors associated with the thermodynamic and dynamic phase transitions. (b) The film thickness dependencies of the critical exponent peak positions ($\varepsilon^{*}$, $\tilde{\varepsilon}^{*}$) for both TPT and DPT. }\label{Fig4}
\end{center}
\end{figure}

In Fig. \ref{Fig3}, the variation of $\tilde{\beta}_{eff}$ is shown as a function of $\tilde{\varepsilon}$ for the DPT in the same film thickness range as for the TPT. In this case, the control parameter is the half-period of the external magnetic field. Figures \ref{Fig2}(b) and \ref{Fig3} display fundamentally and qualitatively similar behavior. Specifically, they illustrate a transition from 3D to 2D behavior that depends on the thickness of the system as well as parameters 
$\varepsilon$ and $\tilde{\varepsilon}$, i.e. the respective distances to the critical point. It is, however, evident from a comparison of Fig. \ref{Fig3} with Fig. 2(b) that the dimensional transition behavior from 3D to 2D criticality is not universal, if one considers the quantitative evolution of the critical exponents $\beta_{eff}$ and $\tilde{\beta}_{eff}$. For instance, the $L_z = 10$ curves in the DPT stay very close to the pure 2D behavior, whereas in the TPT case, this system comes already quite close to 3D type behavior at intermediate distances to the critical point. Thus, quantitatively, the dimensional crossover exhibits significantly different behavior in TPT and DPT. To illustrate this clearly,  we plot the maximum values of the effective critical exponents ($\beta_{eff}^{max}$, $\tilde{\beta}_{eff}^{max}$) as a function of the thickness $L_z$ of the system for both the TPT and DPT in Fig. \ref{Fig4}(a). As evident from the figure, the $\beta_{eff}^{max}$ and  $\tilde{\beta}_{eff}^{max}$ values are strongly dependent on $L_z$. The fundamental similarity between TPT and DPT can be seen in the fact that for $L_z=2$ and $L_z = 20$, the numerical values are indeed very similar. However, for intermediate thicknesses, there is a rather significant deviation in between both, mainly showing that TPT starts to approach 3D behavior already for much thinner films, whereas the DPT preserves almost pure 2D behavior until about $L_z = 10$.  This observation displays the exact same tendencies as earlier experimental findings, where ultrathin ferromagnetic films exhibited 2D-like dynamic behavior while their thermodynamic behavior was found to be closer to the 3D universality class \cite{Quintana3}. Our simulations thus provide a plausible explanation for the coexistence of different critical behaviors within the same physical sample. It is important to underline again that the presence of a maximum in the effective critical exponent as a function of the control parameter is the hallmark of dimensional crossover. In systems with a well-defined crossover from 3D to 2D behavior, the effective exponent initially exhibits an asymptotic behavior towards 3D values as the system approaches the critical point, reflecting the 3D regime. However, upon approaching ever closer to the true critical point, the finite thickness constrains the correlation length, which leads to a transition towards 2D-like scaling and associated critical exponents. This causes the effective exponent to exhibit a peak as the system transitions to the lower-dimensional scaling regime. Thus, both the value and location of this critical exponent peak encode meaningful information about the crossover process: in our case, the maximum peak value reflects the highest degree of 3D-like behavior observed in the system, and its location indicates how far from the critical point the system begins to be influenced by its reduced dimensionality.

To further illustrate the quantitative difference between TPT and DPT dimensional crossover behavior, Fig. \ref{Fig4}(b) displays the variations of the $\varepsilon^*$ and $\tilde{\varepsilon}^*$ values, at which the maximum $\beta_{eff}$ and $\tilde{\beta}_{eff}$ values are obtained, as a function of the film thickness $L_z$, respectively. When comparing the values corresponding to different $L_z$ values, it can be said that $\beta_{eff}$ peaks occur much closer to the critical point in the TPT. In other words, there is a larger $\varepsilon$ range, in which the 3D behavior can build up in the TPT. Therefore, the dimensional crossover is already being observed for smaller $L_z$ values. It is also worthwhile to notice that there is not only an absolute shift in the peak position of the epsilons but also the slopes of $\varepsilon^*$ and $\tilde{\varepsilon^*}$ vs. $L_z$ are different for both cases.

\begin{figure}[h!]
\begin{center}
\includegraphics[scale=1.0]{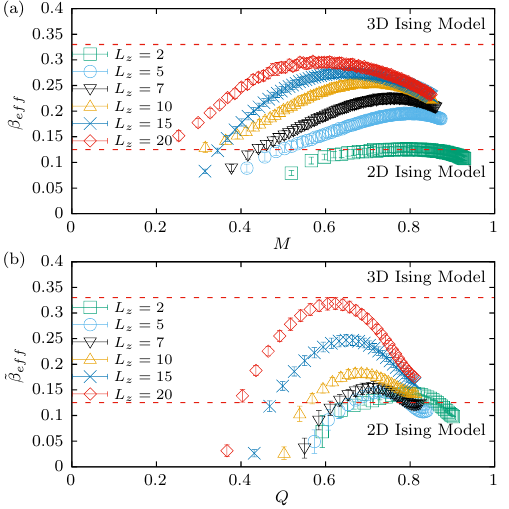}
\caption{Effective critical exponents $(\beta_{eff},\; \tilde{\beta}_{eff})$ as a function of the magnetization $M$ (a) and the dynamic order parameter $Q$ (b). Curves are obtained for different film thicknesses $L_z$ (as indicated) and exhibit the dimensional crossover behavior. The red dashed lines show the critical exponents for both two-dimensional and three-dimensional Ising models.}\label{Fig5}
\end{center}
\end{figure}

To further elucidate the origin of these quantitative differences in between TPT and DPT in terms of their dimensional crossover behavior, the effective critical exponents $\beta_{eff}$ and $\tilde{\beta}_{eff}$ are displayed as a function of
the magnetization $M$  and the dynamic order parameter $Q$ in Figs. \ref{Fig5} (a)-(b) for various values of thickness $L_z$ of the film. When comparing these two figures, we observe that the behavior is not only fundamentally and qualitatively similar, but Fig. \ref{Fig5} also illustrates that the dimensional crossover takes place at fairly comparable values of the order parameter when comparing the TPT and DPT. This observation is particularly intriguing given the very significant differences in the reduced distances to the critical point, represented as $\varepsilon^*$ and $\tilde{\varepsilon}^*$ in Fig.  \ref{Fig4} (b). This suggests that the dimensional crossover point for each particular film, identified by $\beta_{eff}^{max}$ and $\tilde{\beta}_{eff}^{max}$, is rather closely correlated to reaching an intermediate level of the respective order parameter, namely about 60 - 80 per cent of its saturation value. Apparently, these values are achieved already at a much greater relative distance to the critical point in the dynamic case, which then explains the reduced ability of DPT systems to exhibit 3D scaling for intermediate films thicknesses $L_z$. It is very likely that this difference in quantitative behavior is associated with relevantly different surface behaviors in both cases, which is consistent with earlier findings that demonstrated that the surface behavior in the DPT can deviate significantly from that of the TPT \cite{Riego3, Park1}.

\section{Conclusions}
\label{sec:conclusions}
To conclude, this study presents a detailed investigation into the critical behavior and dimensional crossover of DPT in ferromagnetic thin films, using Monte Carlo simulations of the kinetic Ising model. By systematically analyzing the scaling behavior of the dynamic order parameter and comparing it with the TPT, we have illuminated the underlying mechanisms that differentiate these two types of transitions \cite{FSEffect, Data}.  

Our results reveal that while DPT and TPT share commonalities in their scaling laws and critical exponents, the dimensional crossover for the DPT occurs at significantly larger thicknesses than for the TPT. This distinction arises from the relevantly different length-scale and governing principles of the two transitions, with DPT being influenced by time-dependent external fields and non-equilibrium dynamics, whereas TPT adheres to equilibrium-driven processes. The numerical findings reported here align with experimental observations reported in ultrathin Co films \cite{Quintana3}, where dynamic critical exponents were found to be consistent with two-dimensional scaling, whereas thermodynamic critical exponents appeared to follow three-dimensional behavior.  We additionally verified by using a limited set of $L_z$ simulations, that log–log slope analyses of the order parameters  (not shown here) as a function of system sizes reveal different $L_z$ dependencies of the critical exponents for TPT and DPT, which are fully consistent with our main finding in Fig. \ref{Fig4}(a).   Accordingly, our model study captures a plausible mechanism that would produce such a behavior, namely the delayed onset of 3D scaling for the DPT in intermediately thick magnetic films.

As a final note, it should be emphasized that the observed differences in the crossover behavior and effective critical exponents between TPT and DPT may be influenced by several factors beyond dimensionality alone. Possible contributing elements include the nature of surface effects under equilibrium versus nonequilibrium conditions, and the specific dynamical processes, such as relaxation times and nucleation dynamics,  involved in each case. While our present study focuses on characterizing the basic phenomenology of this difference, exploring how these factors quantitatively impact the scaling behavior (particularly whether the exponent difference can be systematically enhanced or reduced) remains an open and intriguing direction for future research.
\begin{acknowledgments}
The numerical calculations reported in this paper were performed at T\"{U}B\.{I}TAK ULAKBIM (Turkish agency), High Performance and Grid Computing Center (TRUBA Resources). Work at nanoGUNE was supported by the Spanish Ministry of Science and Innovation under the Maria de Maeztu Units of Excellence Program (Grant No. CEX2020-001038-M) and Project No. PID2021-123943NB-I00 (OPTOMETAMAG).
\end{acknowledgments}


\begin{thebibliography}{100}
\bibitem{FSEffect} To assess the potential influence of lateral finite-size effects on our results, we performed additional test simulations with a lateral dimension of 
$L=256$, using the same film thickness values 
$L_z$ as in the present study. These simulations show that the effective critical exponent curves for both the thermodynamic and dynamic cases remain essentially unchanged when increasing 
$L$ from 256 to 512, indicating that the chosen system sizes are sufficiently large to suppress finite-size effects in the lateral directions.
\bibitem{Data} The data are available upon reasonable request from the authors.
\end{thebibliography}
\end{document}